\documentclass[fleqn,usenatbib]{mnras}

\usepackage{newtxtext,newtxmath}

\usepackage[T1]{fontenc}
\DeclareRobustCommand{\VAN}[3]{#2}
\let\VANthebibliography\thebibliography
\def\thebibliography{\DeclareRobustCommand{\VAN}[3]{##3}\VANthebibliography}

\usepackage{graphicx}	
\usepackage{amsmath}	

\title[Accretion disc midplane backflow]{Accretion disc backflow in resistive MHD simulations}

\author[R. Mishra et al.]{
R. Mishra,$^{1}$\thanks{E-mail:rmishra@camk.edu.pl}
M. Čemeljić,$^{1,2,3}$
W. Kluźniak$^{1}$
\\
$^{1}$Nicolaus Copernicus Astronomical Center of the Polish Academy of Sciences, Bartycka 18, 00-716, Warsaw, Poland\\
$^{2}$Academia Sinica Institute of Astronomy and Astrophysics, P.O. Box 23-141, Taipei 106, Taiwan\\
$^{3}$Research Centre for Computational Physics and Data Processing, Silesian University in Opava, Bezru\v{c}ovo n\'am.~13, CZ-746\,01 Opava, Czech Republic
}

\date{Accepted XXX. Received YYY; in original form ZZZ}

\pubyear{2015}

\begin{document}
\label{firstpage}
\pagerange{\pageref{firstpage}--\pageref{lastpage}}
\maketitle

\begin{abstract}
  We investigate accretion onto a central star, with the size, rotation rate, and magnetic dipole of a young stellar object, to study the flow pattern (velocity and density) of the fluid within and outside of the disc. We perform resistive MHD simulations of thin $\alpha$-discs, varying the parameters such as the stellar rotation rate and magnetic field, and (anomalous) coefficients of viscosity and resistivity in the disc. To provide a benchmark for the results and to compare with known analytic results, we also perform purely hydrodynamic simulations (HD) for the same problem. Although obtained for a different situation with differing inner boundary condition, the disc structure in the HD simulations closely follows the analytic solution of \citet{KK00}---in particular a region of ``midplane'' backflow exists in the right range of radii, depending on the viscosity parameter. In the MHD solutions, whenever the magnetic Prandtl number does not exceed a certain critical value, the midplane backflow  exists throughout the accretion disc, extending all the way down to the inner transition zone where the disc transitions to a magnetic funnel flow. For values of the magnetic Prandtl number close to the critical value the backflow  and the inner disc undergo a quasiperiodic radial oscillation, otherwise the backflow is steady, as is the disc solution. From our results, supplemented by our reading of the literature, we conclude that midplane backflow is a real feature of at least some accretion discs, whether HD $\alpha$-discs or MHD discs, including ones driven by MRI turbulence.
\end{abstract}



\section{Introduction}
\label{intro}
Mass transfer onto massive objects occurs through gaseous accretion discs in most astrophysical scenarios. Accretion in the disc is driven  by the viscous stresses that arise from the differential rotation of  the disc fluid. These stresses  help in transporting matter inwards and angular momentum outwards. 
In the classical $\alpha$-disc model of \citet{SS73} the dominant component of viscous stress tensor  is proportional to the pressure: $T_{{r\phi}} = \alpha P$. This results in the kinematic viscosity being parameterized by the  dimensionless parameter $\alpha$ such that $\nu \sim \alpha c_\mathrm{s} H$, where $c_\mathrm{s}$ is the sound speed and $H$ is the disc thickness.
In such one dimensional approach to a steady accretion disc, with a constant mass accretion rate, the radial motion of gas is purely inward \citet{SS73,Pringle}. However, in 2D or 3D, while the density-weighted vertical average of the radial velocity ($\varv_{r}$) is negative, i.e. towards the central object, the flow need not be in the same direction everywhere---under some conditions a part of the disc flow is found to be in the opposite direction, away from the central object. This is referred to as a ``backflow'' in the accretion disc, and it usually appears in the disc midplane as a consequence of dynamical response to viscous torque. To distinguish from the outflows which might occur near the disc surface and backflow in case of AGN jets in the work by \citet{AGNbackflow}, we introduce the term ``midplane backflow.'' 
\\
  \begin{figure}
   \centering
   \includegraphics[width=0.9\hsize]{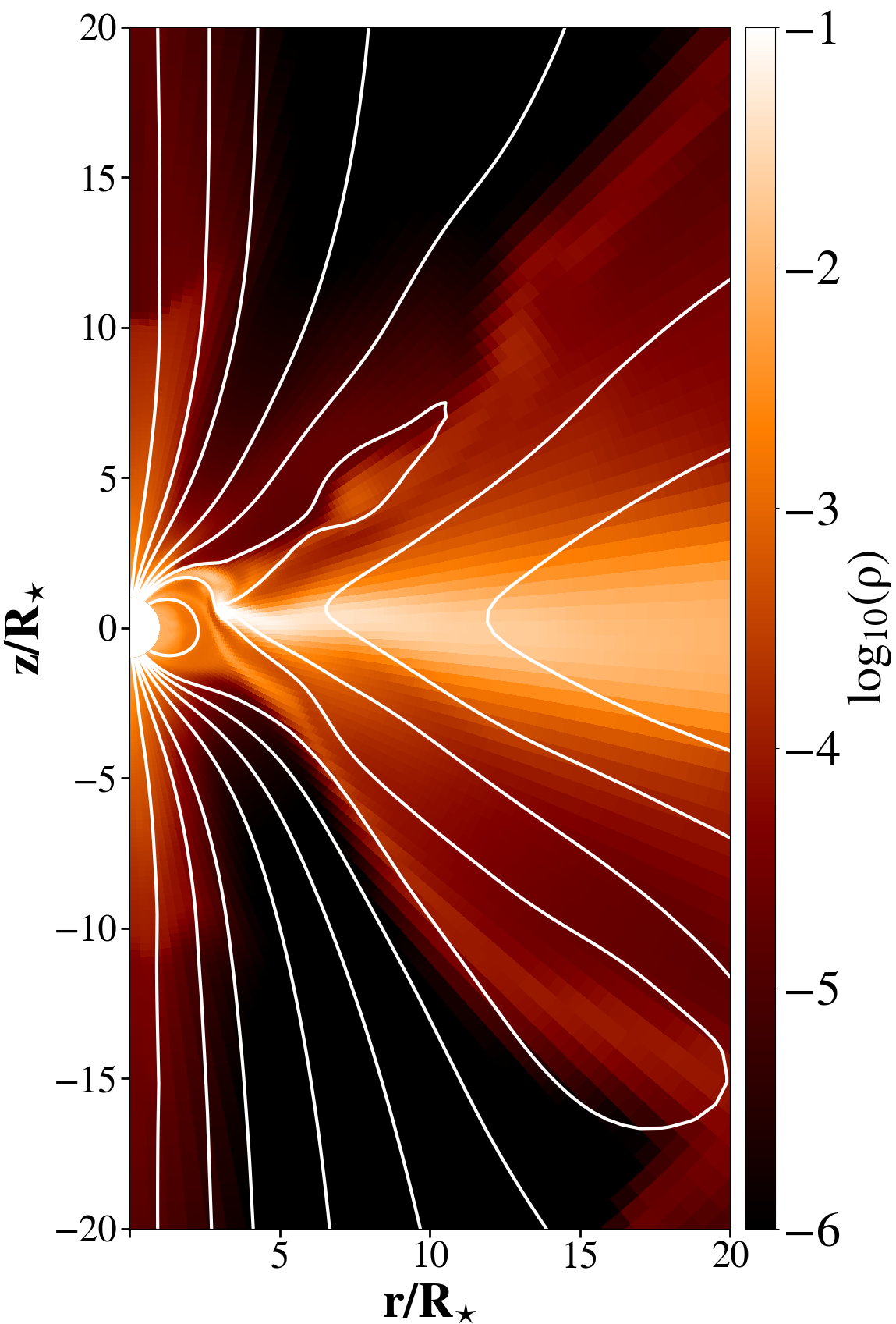}
      \caption{Density distribution, shown in a logarithmic color grading, in the star-disc system in one of our 2D axisymmetric simulations including a complete meridional half-plane. White lines indicate the poloidal magnetic field lines. The central object is surrounded by an accretion disc, and from the magnetosphere of such a star-disc system various outflows are launched. Accretion onto the star proceeds through a funnel, typically  switching with time from one to another stellar hemisphere.  In the disc itself, accretion flow is mainly directed towards the star, but a part of it can be redirected away from the star.}
         \label{simforsce}
   \end{figure} 
 
  
  \begin{figure}
   \centering
   \includegraphics[width=0.9\hsize]{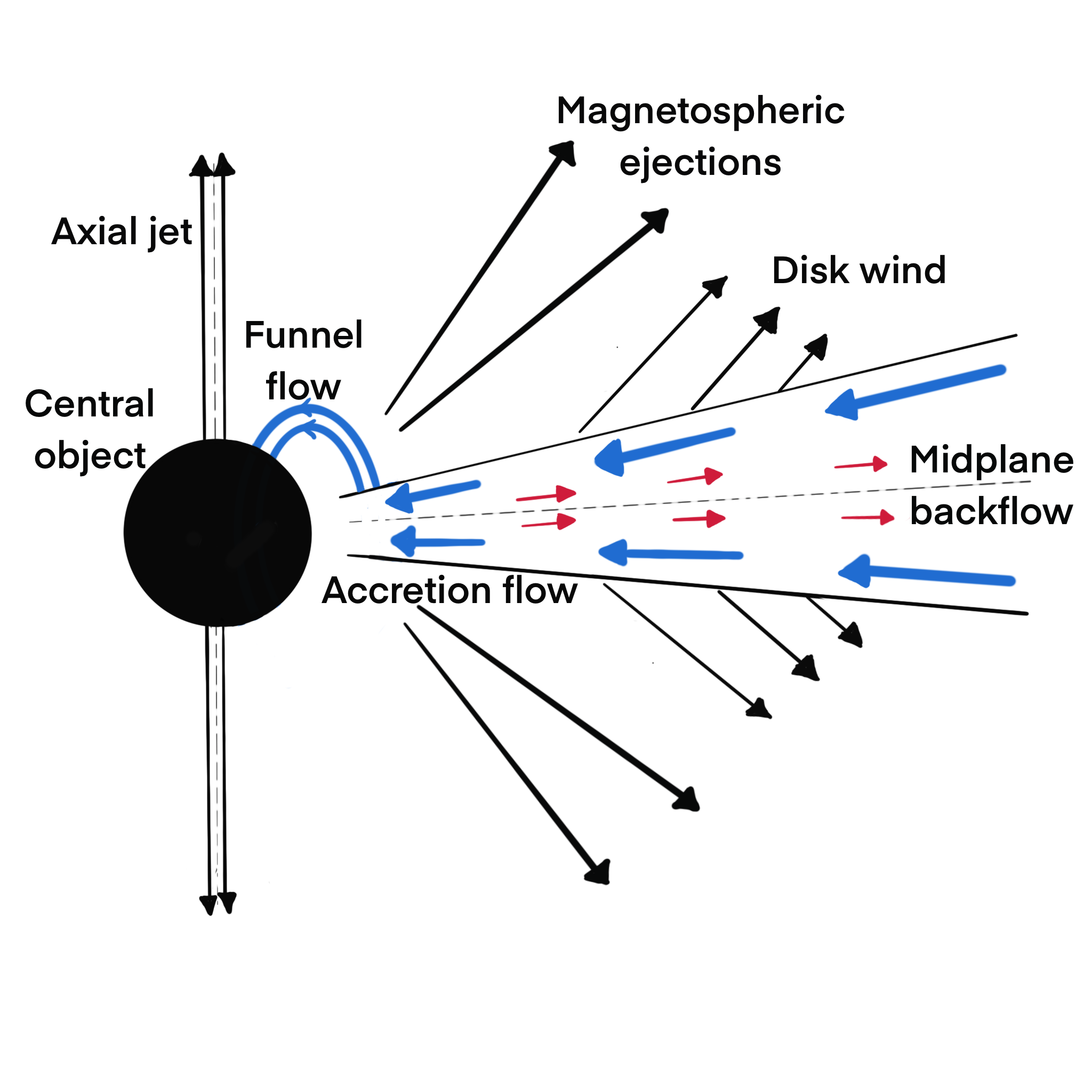}
      \caption{Schematic diagram representing the flows in the star-disc system shown in Fig.~\ref{simforsce}. The central object is surrounded by an accretion disc. Outside the disc there are outflows marked by the black arrows such as: axial outflow, magnetospheric ejections and disc wind. Inside the disc, the main component (marked by blue arrows) is accretion flow towards the central object. Close to the inner edge of the disc, matter can be lifted above the disc midplane and channeled onto the star through the funnel flow, following the magnetic field lines. In some cases there is a midplane backflow in the disc, marked by red arrows.}
         \label{Schematic}
   \end{figure} 

   Disc backflow was discovered by \citet{urp84a}, who analytically obtained two  dimensional  approximate  structure in $r-z$ plane  of an accretion disc segment using the Shakura \& Sunyaev alpha prescription. He found an outflow in the equatorial plane and inflow near the surface of the disc for all values of viscosity parameter in his local calculations. A similar result was obtained subsequently in the numerical work of \citet{Ane}, although the result was ascribed to meridional circulation. The same term was used also by  \citet{kl92}, in what is now recognized---with hindsight---as a paper presenting the first ever numerical simulation of backflow, the vertical circulation pattern being an artefact of their outer boundary conditions (inflow at large radii across the height of the disc).
 \citet{roz94}, found backflow extending throughout the disc in their numerical work, as they did not enforce inflow at the outer boundary. 
 
 Global analytic solutions in 3D were obtained by \citet{phdthesiskita} 
 and \citet{KK00} (hereafter KK) using the polytropic pressure--density relation, and including a zero--torque inner boundary condition. For consistency, just as in the pioneering work of Urpin, all components of the viscous stress tensor were included  (a crucial step for obtaining backflow). In the KK axisymmetric calculations, backflow was obtained in the disc-mid plane for alpha viscosity less than a critical value. Accretion takes place along the entire height of the disc for $\alpha$ larger than the critical value, but only at higher altitudes for lower values of the viscosity parameter. These results were confirmed by \citet{2002RG} for a non-polytropic equation of state with slightly different assumptions, including the treatment of thermal effects which were excluded in KK.
 
 Some more recent examples of numerical simulations with backflow include simulations of $\alpha$--discs by  \citet{igu},  \citet{Lee}. The results are typically in agreement with the KK solution, although for some simulations backflow occurs in the surface layers, and not in midplane region.
 \citet{TakuLin} explain ``the outward flow at the midplane'' as a result of the shallower gradient of density there (as compared with the surface layers).
 \citet{Phili} suggest that  direction of the flow at the midplane is intimately connected to the behavior of the vertical stress, $T_{z\phi}$.
 
The midplane backflow has often not been thought important in the simulations. In the work of \citet{ZF09} backflow was also obtained in the simulations, but it was disregarded as  ``certainly unphysical and arising only from the functional form of the stress tensor used to mimic turbulence''. In the work of \citet{Pan}, the $\tau_{\theta \phi} $ component of the viscous stress tensor was intentionally neglected, in order to avoid the backflow along the disc midplane.
However, midplane backflow appears also in simulations where accretion is driven by magneto-rotational  instability (MRI) in \citet{bm}, global  simulations in ideal magneto-hydrodynamics (MHD) by \citet{Zhustone} and general-relativistic radiative magneto-hydrodynamics (GRMHD) simulations of  radiatively inefficient accretion flows by \citet{white}.
This suggests the reality of backflow in a variety of accretion discs.

 
 In the present paper, we simulate thin accretion disc around young stellar objects (YSOs) in resistive MHD in order to study backflow in the presence of a stellar magnetic field as a function of viscosity. This can be also rescaled for other objects with similar geometry. A snapshot of our star-disc simulation has been presented in Fig.~\ref{simforsce}. In Fig.~\ref{Schematic}, we present a schematic diagram presenting different flows in a star disc system, corresponding to our simulation results.
 
 The paper is organized as follows: In \S 2  we briefly review some important aspects of analytical solution with backflow from KK. In \S 3 we provide background of numerical set up. In \S 4  we present results from purely hydrodynamical (HD) cases, as well as MHD ones. In \S 5 we discuss observational significance of the work, and shed light on backflow in the context of MRI dominated disc. In the final section we summarize the conclusions of this work.
 
  \begin{figure*}
   \centering
  
   \includegraphics[width=\hsize]{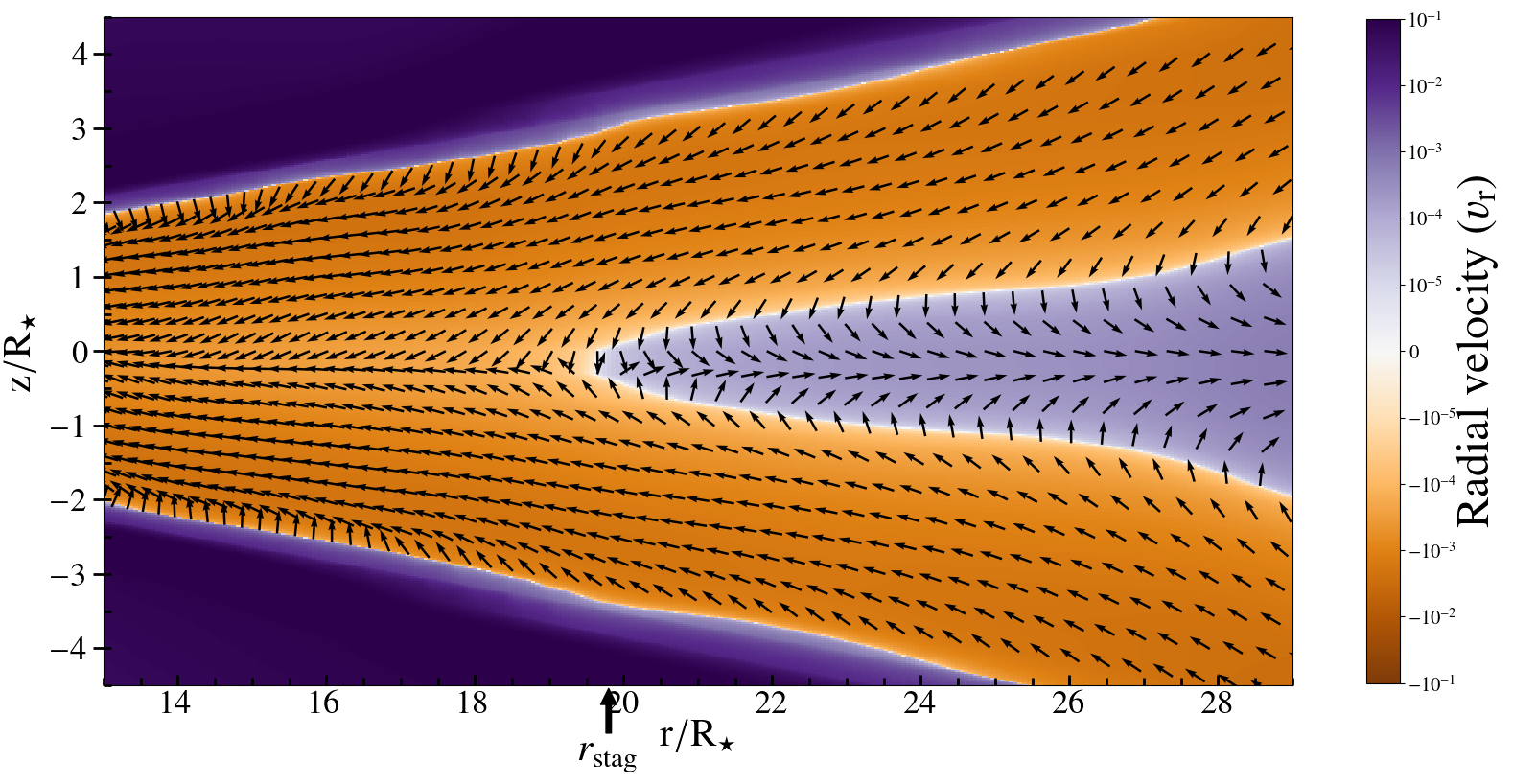}
  
   \caption{ A close up view into a part of accretion disc in a quasi-stationary state in our purely hydrodynamic simulation with the viscous coefficient $\alpha=0.4$. The resolution is $r\times\theta=[217\times200]$ grid cells, in the meridional half-plane $\{(r,\theta)\}=(1,29)\times (0,\pi)$. The disc midplane is positioned at $\theta=\pi/2$. The arrows represent the direction of flow.  Analytical solutions in KK show that with such a viscosity a midplane backflow occurs in the disc, in which material flows away from the central object. We confirm the predicted steady backflow in our simulation. The light violet region in the accretion disc indicates the backflow region. For this simulation the stagnation radius $r_{\mathrm{stag}}=19.8\,R_*$, indicates that the circle in the midplane, outside which the accretion flow is away from the star and within which it is towards the star, is positioned at 19.8 stellar radii.}
   \label{fullpihd}
\end{figure*}

 \section{Klu\'zniak-Kita disc solution with backflow}
To provide context for our MHD results we start with a summary of the analytic results for a hydrodynamic steady, thin alpha disc.
   \citet{KK00} presented a global three dimensional solution of accretion disc. It is valid for any polytropic\footnote{A polytropic equation of state guarantees isentropy, in conformity with the thin disc approximation that the dissipated heat is radiated locally.} viscous thin Keplerian accretion disc, and has been  obtained by an asymptotic expansion in a small parameter $\epsilon=\tilde H/\tilde R$, which is the disc aspect ratio. In their work an inner boundary condition has been introduced  with a vanishing viscous torque at a certain radius corresponding to the disc inner edge. They have also adapted the alpha prescription by incorporating the $z$ dependence in the kinematic viscosity. One of the main results was that beyond a certain distance along the midplane of the disc there may or may not be backflow, depending on the $\alpha$ parameter.
     From their solution, the equation for radial velocity in the equatorial plane (midplane of the disc) is given by,
   \begin{equation}
       \varv_\mathrm{r}(r,0)= - \alpha\epsilon^{2}\Omega_{\mathrm{K} \star} \tilde{R} \bigg(\frac{h^2}{r^{5/2}}\bigg)\bigg[2\bigg(\frac{d \ln h}{d\ln r}\bigg)-\Lambda\bigg(1+\frac{32}{15}\alpha^2\bigg)\bigg]
   \end{equation}
   where,
   \begin{equation*}
       \Lambda =\frac{11}{5}\left(1+ \frac{64}{25}\alpha^2\right)^{-1}\ .
   \end{equation*}
Here $\alpha$ follows the \citet{SS73} alpha prescription,
and $h=H/r$ is the dimensionless disc height. All the radial distances are scaled by  some characteristic radius $\tilde{R}$. The Keplerian angular velocity is scaled by $(GM_\star/\tilde{R}^3)^{1/2}\equiv \Omega_{\mathrm{K}\star}$. At smaller radii the above equation shows that  $\varv_\mathrm{r}(r, 0) < 0$, i.e there is always inflow. But at larger radii, for values of $\alpha$ below a certain value, it results in positive radial velocity  $\varv_\mathrm{r}(r, 0) > 0$, which indicates outflow in the equatorial plane away from the central accretor. The overall pattern is similar to the one found in our HD simulation (Fig.~\ref{fullpihd}). On the other hand, when $\alpha$ is close to unity, then $\varv_\mathrm{r}(r, 0) < 0$ everywhere, which indicates the equatorial flow is directed inwards towards the central star for all radii.
 The critical value above which there is no backflow is
   \begin{equation}
       \alpha_{\mathrm {cr}}\approx0.685.
       \label{alfcrit}
   \end{equation}
 When $\alpha < \alpha_{\mathrm{cr}}$ midplane backflow continues all the way to the outer boundary, where the disc terminates (or begins). 
 The radius for which the radial velocity is zero in the disc midplane, i.e the terminus of the backflow region, is the stagnation radius $r_{\rm stag}$. The equation for stagnation radius in KK is given by:
 
     \begin{equation}
       \frac{r_{\mathrm {stag}}(\alpha)}{r_+}= \left\{
       \frac{1+6\big[\Lambda\big(1+\frac{32}{15}\alpha^2\big)-2\big]}{6\big[\Lambda\big(1+\frac{32}{15}\alpha^2\big)-2\big]}\right\}^2.
   \end{equation}
Here $ r_+ $ is a natural length scale introduced into the problem via the inner boundary condition (vanishing torque at $ r_+ $). The stagnation radius is a function of viscous coefficient $\alpha$. As $\alpha$ increases, stagnation radius increases. When $\alpha$ approaches the critical value $\alpha_{\mathrm {cr}}$, the stagnation radius becomes infinite---there is no backflow for $\alpha > \alpha_{\mathrm{cr}}$. KK concluded that for small values of $\alpha$, the accretion flow turns around and feeds a backflow close to the equatorial plane of the disc, transporting fluid outwards to the outer boundary of the accretion disc. They consider midplane backflow to be a general feature of a geometrically thin accretion disc.

 \begin{figure*}
\centering
\includegraphics[width=1.0\columnwidth]{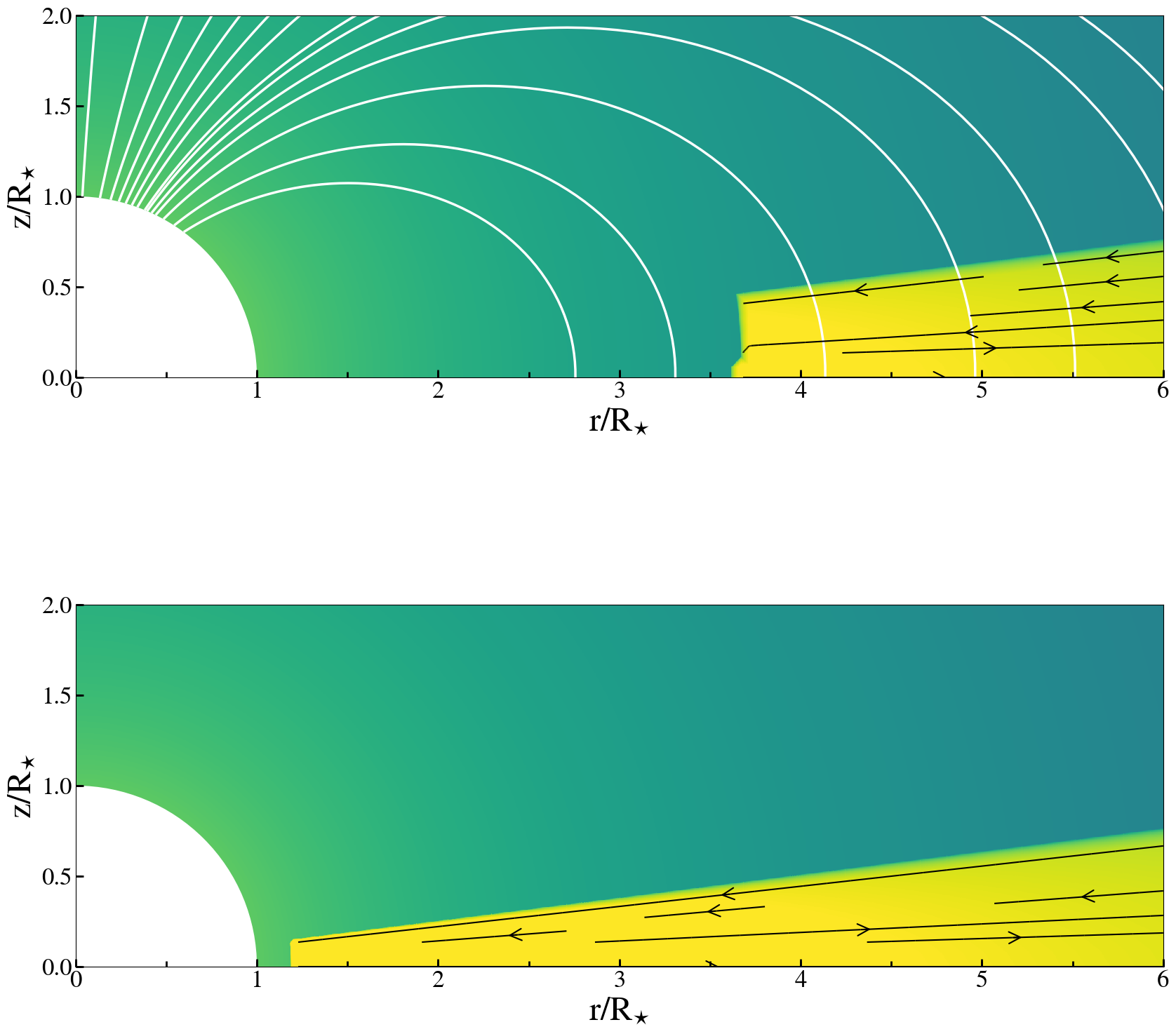}
\includegraphics[width=1.0\columnwidth]{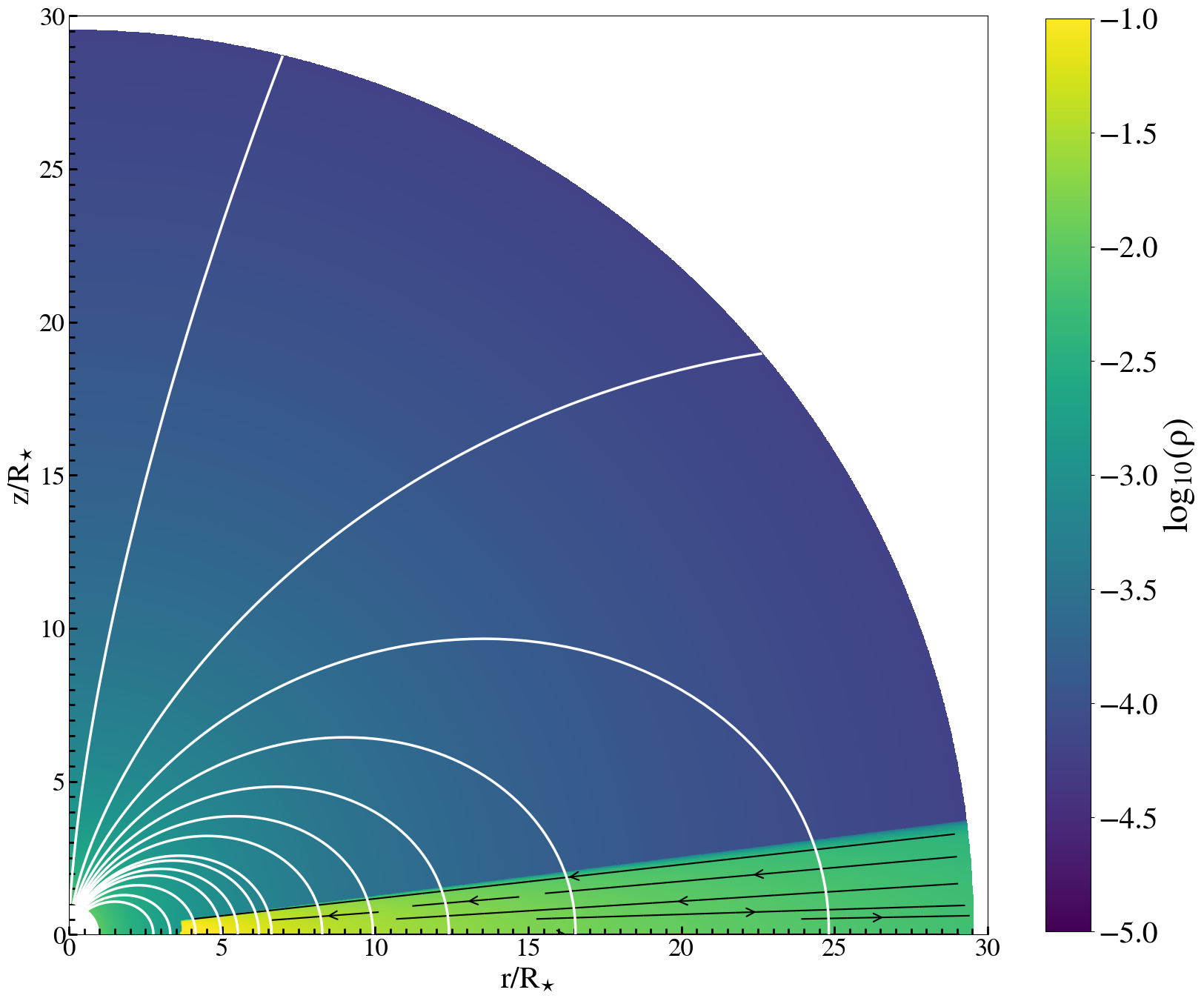}
  
\caption{The initial, $t=0$, configuration in viscous purely hydrodynamic ({\it bottom left panel}) and magnetic ({\it top left panel}) disc. Density distribution is shown with logarithmic color grading, and solid black lines with arrows show poloidal velocity streamlines. The complete domain is shown in the magnetic case ({\it right panel}), with the dipolar magnetic field lines shown with white solid lines. The resolution in such simulations in our work is $r\times\theta=[217\times100]$ grid cells, in a quadrant of the meridional plane $\{(r,\theta)\}=(1,29)\times (0,\pi/2)$.}
         \label{initial}
\end{figure*}   

\begin{figure*}
\centering
\includegraphics[width=1.6\columnwidth]{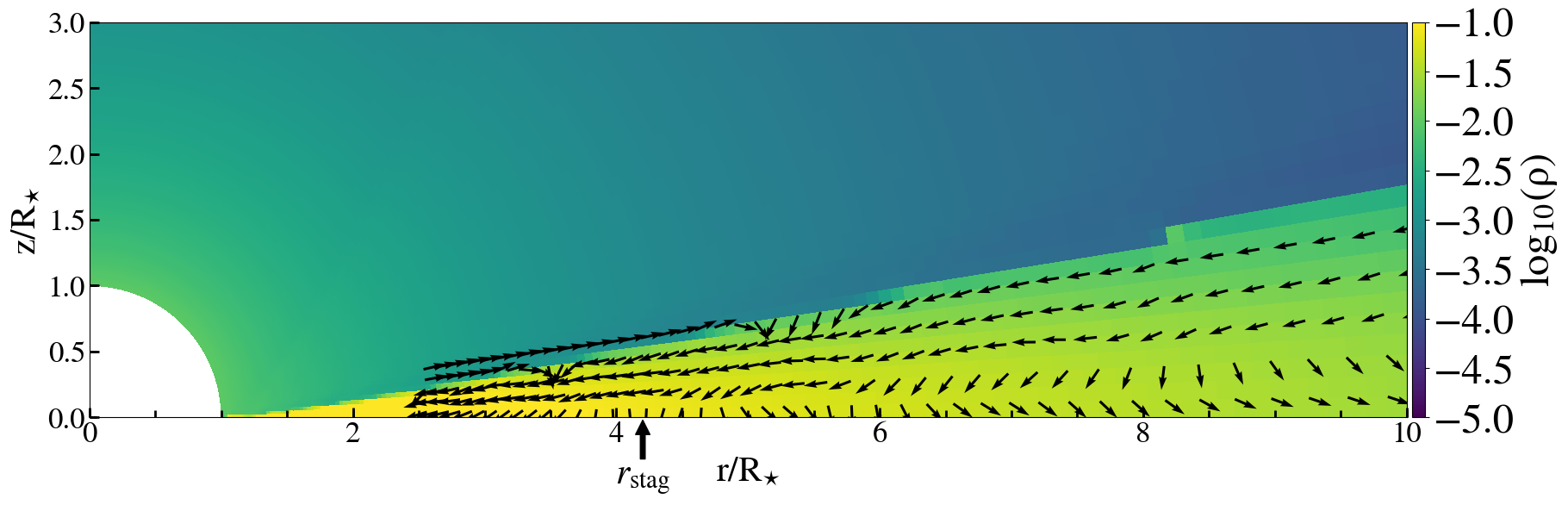}\\
\includegraphics[width=1.6\columnwidth]{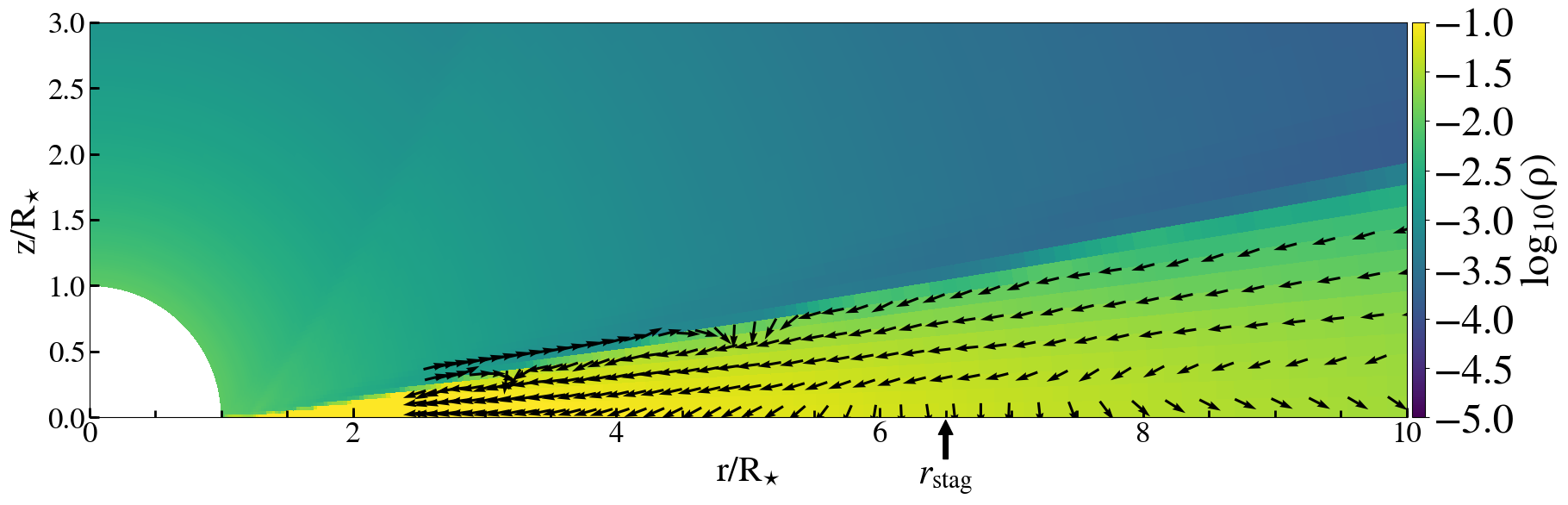}\\
\includegraphics[width=1.6\columnwidth]{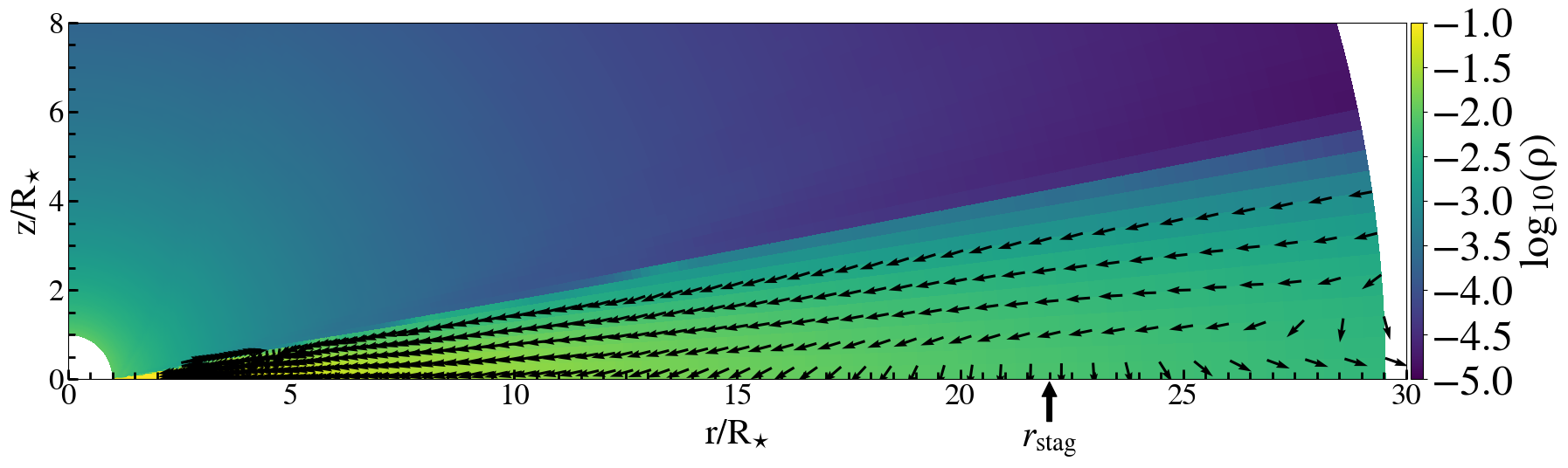}\\
\includegraphics[width=1.6\columnwidth]{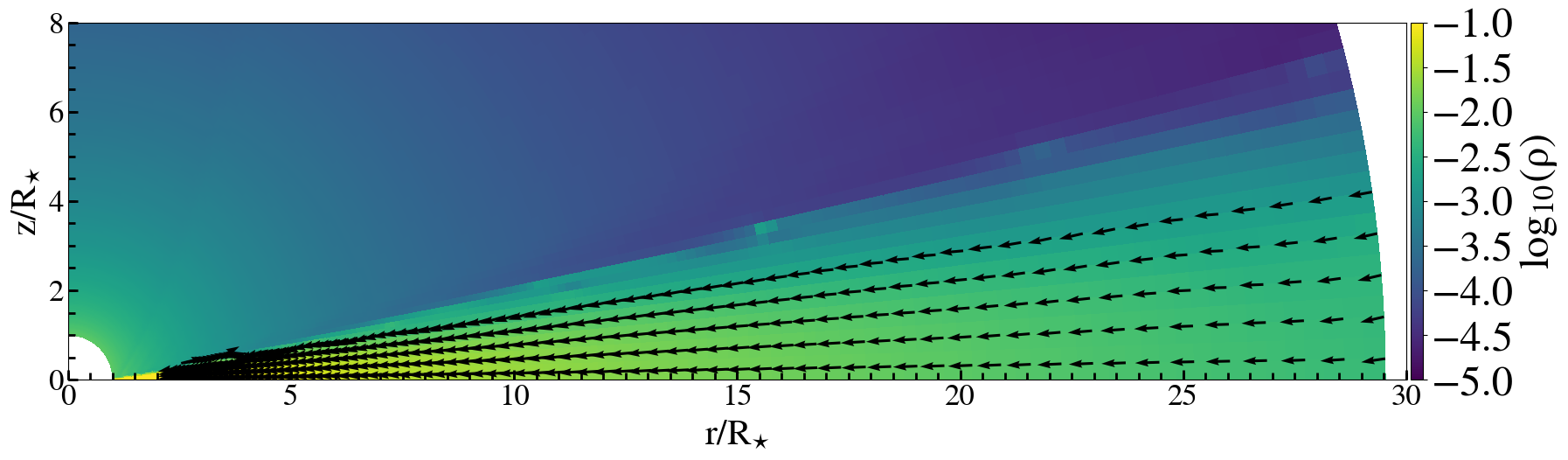}\\
\caption{HD simulations with $\alpha_\mathrm{v}$=0.1, 0.2, 0.4 and 1.0, top to
bottom panels, respectively. The density is shown in logarithmic grading. Black arrows indicate velocity vectors in the disc. The position of stagnation radius is marked as $r_\mathrm{stag}$. In the bottom simulation there is no backflow, as $r_\mathrm{stag}\rightarrow\infty$.}
 \label{stag}
\end{figure*} 

    \begin{figure*}
   \centering
   \includegraphics[width=1.0\columnwidth]{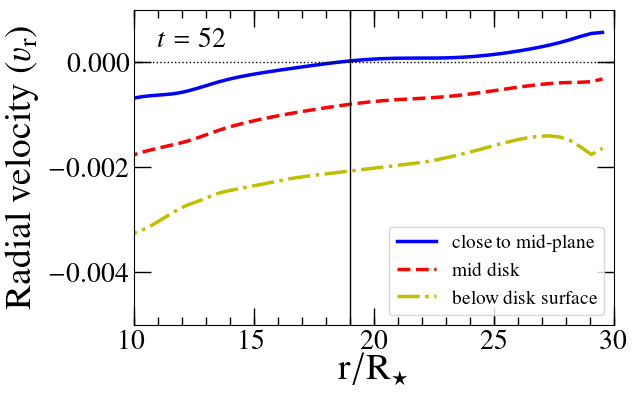}
   \includegraphics[width=1.0\columnwidth]{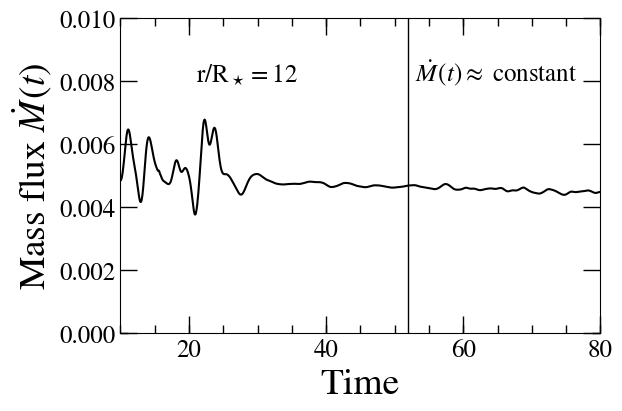}
   \includegraphics[width=\hsize]{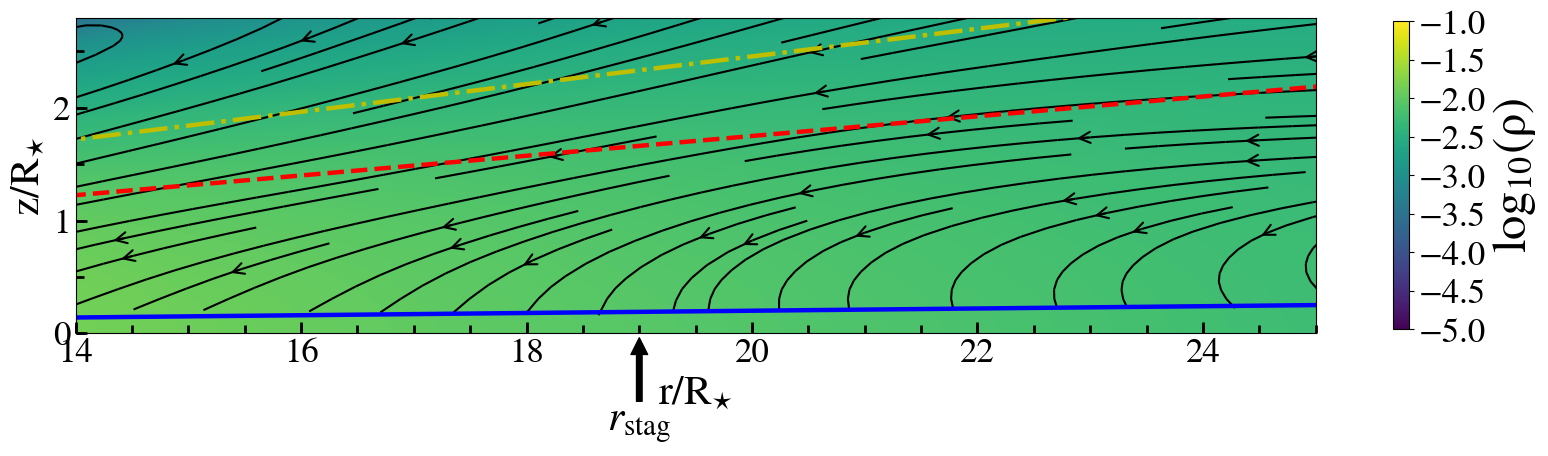}
   \caption{HD simulation with $\alpha_{\mathrm v}=0.4 $ and $\Omega = 0.2\,\Omega_{\rm br} $ with the disc in a quasi-stationary state. {\it Top left panel:} Radial velocity component as a function of radius across different heights of the disc as marked by lines of the same color and style in the bottom panel. Stagnation radius, where the midplane radial velocity component changes direction, is marked with the vertical solid line. {\it Top right panel:} The mass flux in the disc at $r/R_\star$ = 12, as a function of time. The vertical line corresponds to the snapshot presented in the bottom panel, taken during the quasi-stationary state. {\it Bottom panel:}  Snapshot of the meridional plane at $t=52$. The density is shown in a logarithmic color grading. The stream lines represent the flow velocity. The stagnation radius is at  $r_{\mathrm{stag}}= 19R_\star$.
   Streamlines close to the equatorial midplane (below the blue line) have been suppressed for clarity.}
         \label{hd}
\end{figure*}

    \begin{figure}
     \includegraphics[width=0.99\hsize]{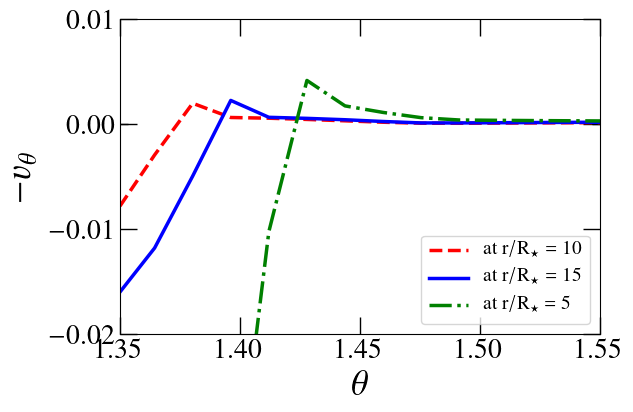}
      \caption{"Vertical" ($-\varv_\theta$) component of velocity in quasi-stationary state (at $t=198$) as a function of $\theta$ ($\theta= \pi/2$ corresponds to the midplane of the accretion disc) in the HD simulation with $\alpha_{\mathrm v}=0.1 $ and $\Omega = 0.1\,\Omega_{\rm br}$. In the backflow region, the vertical velocity is positive close to the midplane, i.e., the flow diverges from the midplane. Here, the stagnation radius is at $r_{\mathrm{stag}}=4.0\,R_\star$. }
         \label{vtheta}
   \end{figure} 

\begin{figure*}
   \centering
   \includegraphics[width=1\columnwidth]{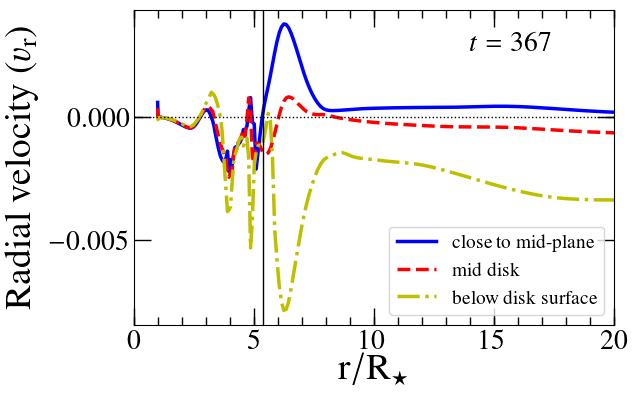}
   \includegraphics[width=1.0\columnwidth]{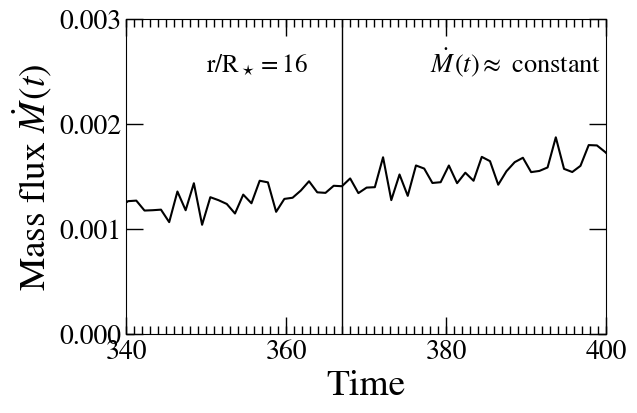}
   \includegraphics[width=\hsize]{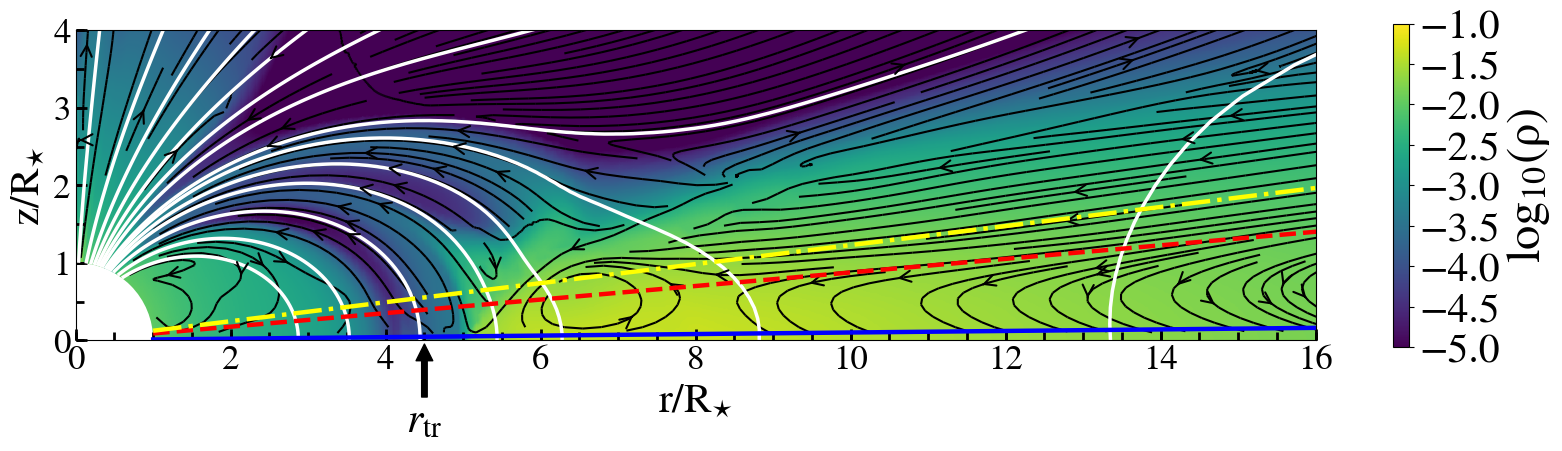}
   \includegraphics[width=\hsize]{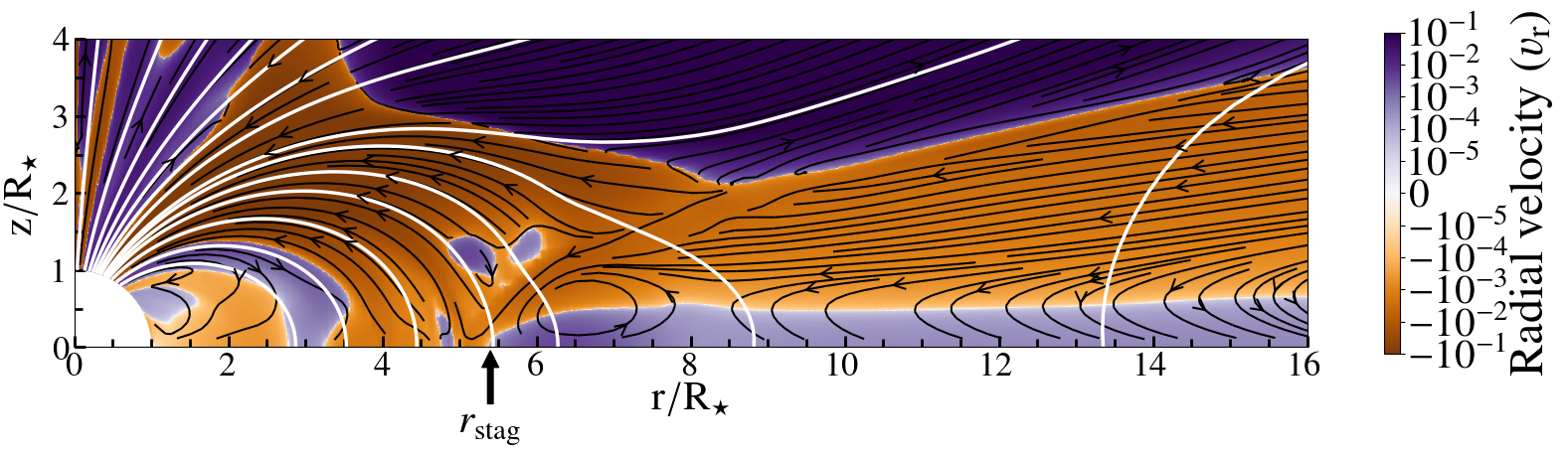}
      \caption{MHD simulation with $\alpha_{\mathrm v}=0.4$, $\alpha_{\mathrm m}=1$, with the initial dipolar magnetic field of $1000\,$G, in the case of a star rotating at $0.1\,\Omega_\mathrm{br}$  (corotation radius $r_\mathrm{co}=4.6\,R_\star$). {\it Top left panel:} The radial velocity component shown as a function of radius across different heights of the disc as depicted in the middle panel. Stagnation radius is marked with the vertical solid line. {\it Top right panel:} Mass flux through the disc at  $r/R_\star = 16$ shown as a function of time. Changes during the chosen interval are small, so we consider it as a quasi-stationary state. The vertical solid line corresponds to the time of the snapshot presented in the middle panel. {\it Middle panel:} Snapshot of the meridional plane at $t =367$. The density in a logarithmic color grading. The stream lines represent the flow velocity (suppressed for clarity near the midplane). The disc truncation radius is at $r_{\mathrm{tr}}=4.5R_\star$ . The style and color of the indicated lines of constant $\theta$ correspond to the cuts shown in the top left panel. {\it Bottom panel:} Snapshot of the meridional plane at $t =367$. Radial velocity in a logarithmic color grading. The violet region near the disc midplane, outwards from the stagnation radius marked at 5.4 stellar radii, indicates backflow.}
         \label{Mhd}
   \end{figure*}
   \begin{figure}

   \includegraphics[width=\hsize]{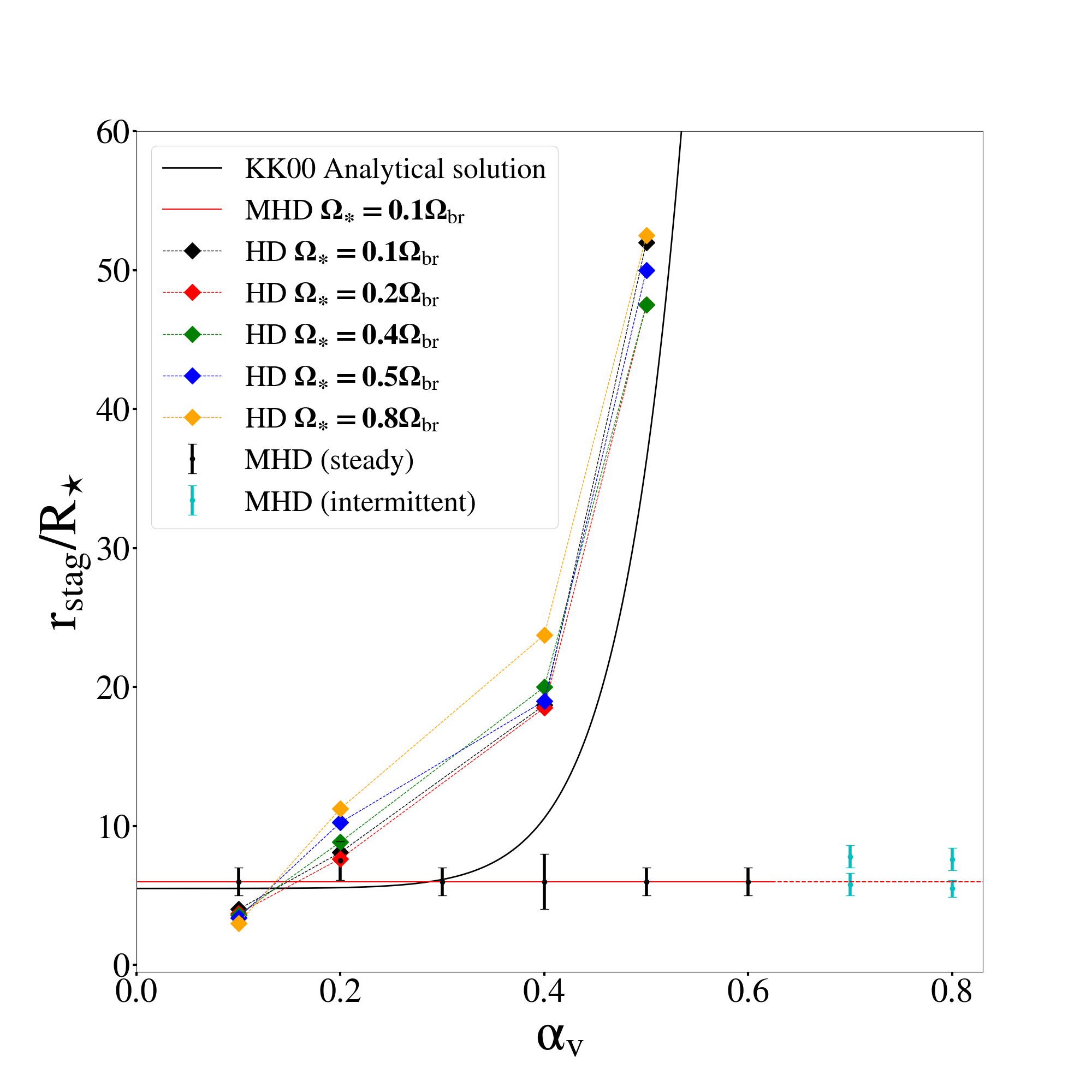}
      \caption{Position of the stagnation radius in HD and MHD simulations with the different viscosity coefficients. In HD the curves have been presented for simulations with various stellar rotations ranging from $0.1\,\Omega_\mathrm{br}$ (slowly rotating stars) to $0.8 \,\Omega_\mathrm{br}$ (fast rotation). For the purpose of comparison we also add the black dashed curve from the \citet{KK00} purely HD analytical solution. In MHD (runs with a $\sim 1.4 \times 10^{36}\,\mathrm{G\,cm^3}$ stellar dipole and $\Omega_\star = 0.1\,\Omega_\mathrm{br}$) we mark the stagnation radius for the cases of steady backflow (black data points, with error bars) and intermittent one (blue data points marking the range of variation of $r_\mathrm{stag}$). In the steady MHD case the stagnation radii can be fitted with a constant (solid red line).}
         \label{FigVibStab}
   \end{figure} 
\begin{figure}
   \centering
  
   \includegraphics[width=1.0\columnwidth]{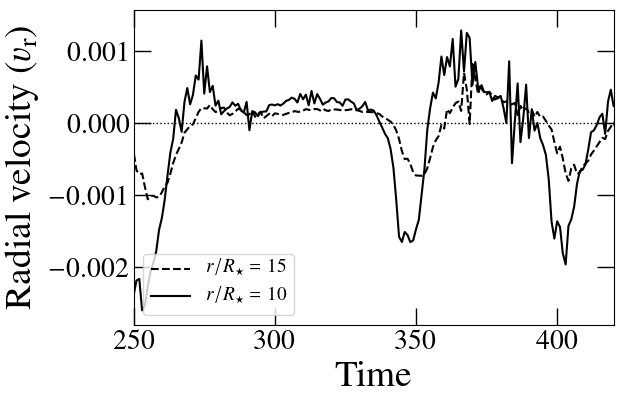}
  
   \caption{Radial velocity calculated at $r/R_{\star} = 10$ close to the midplane of accretion disc as a function of time from the MHD simulation with  $\alpha_{\mathrm {v}}= 0.8 $, $\alpha_{\mathrm {m}}= 1 $, $B_\star=1000\,$G and $\Omega = 0.1\,\Omega_{\rm br}$ with intermittent backflow, which is present in the intervals with positive radial velocity. The quasi-period is a few tens of orbital periods at this radius.}
   \label{int}
\end{figure}

  
%

\section{Simulations setup}\label{setup}
 We performed star-disc simulations using the setup presented in detail in \citet{cem19}. In the  initial conditions we set a thin, viscous, purely hydrodynamical KK disc around YSO with $M_\star = 0.5 M_\odot$ and $ R_\star = 2 R_\odot$. For the MHD case, we add a stellar dipolar magnetic field and include anomalous (much larger than computed from the microscopic considerations) Ohmic resistivity in the disc. Disc resistivity is described with resistivity parameter $\alpha_{\mathrm{m}}$. The simulations are done in 2.5D: two-dimensional and axisymmetric, with the computation of the time evolution of the velocity and magnetic field in $\varphi$ direction. A logarithmically increasing grid in radial direction in spherical coordinates is used, with uniformly spaced colatitudinal grid. For all simulations resolution is $r\times\theta=[217\times100]$ grid cells, in a quadrant of the meridional plane $\{(r,\theta)\}=(1,29)\times (0,\pi/2)$ with the exception of one simulation where resolution is $r\times\theta=[217\times200]$ grid cells, in the meridional  half-plane $\{(r,\theta)\}=(1,29)\times (0,\pi)$, with $r$ in units of $R_*$.
For initial conditions we use the HD $\alpha$-disc solution of \citet{KK00}. In Fig.~\ref{initial} we present the initial setup from HD and MHD simulations. We use the publicly available PLUTO code (v.4.1) \citep{m07} for our simulations.  In the disc outer radial boundary  we feed the disc continuously with the initial hydrodynamical values corresponding to the KK solution with the appropriate value of $\alpha_\mathrm{v}$.

\begin{figure*}
\centering
\includegraphics[width=1.0\columnwidth]{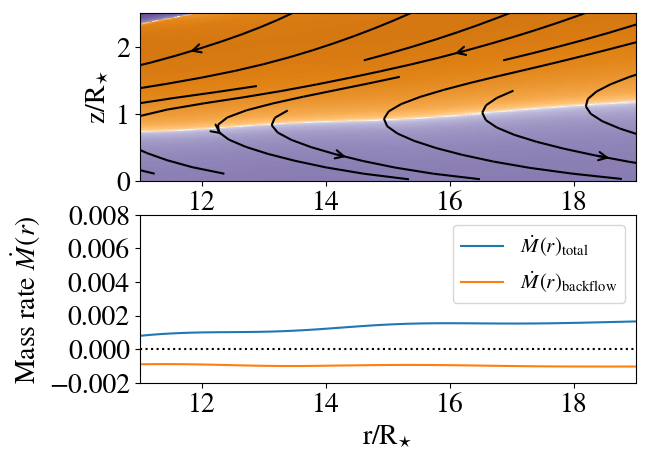}
\includegraphics[width=1.0\columnwidth]{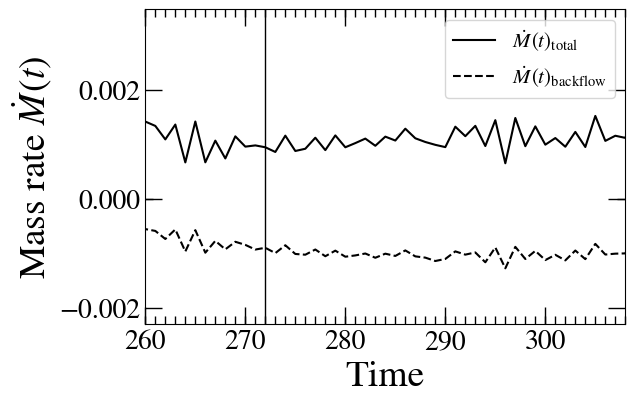}
\caption{MHD simulation with $\alpha_{\mathrm m}=1$, $\alpha_{\mathrm v}=0.5$, $B_\star=1000\,$G and $\Omega = 0.1\,\Omega_{\rm br}$  {\it Left panel:} The net mass accretion rate, and the mass flow rate in the backflow (with the sign convention of Eq.~\ref{massflow}) as a function of radius, calculated in the quasi-stationary state at $t=272$ as marked by vertical line in the right panel. The violet part of the top panel indicates the backflow region. {\it Right panel}: Total mass flux and backflow  mass flux calculated at $r/R_\star = 15$ as a function of time. For $t>272$ (to the right of the vertical line) the flow can be treated as steady.
\label{massflux}}
\end{figure*}  
Pluto solves the set of MHD equations including the viscous and resistive terms. In the HD setup the magnetic part is omitted. The equations in the cgs units are given by:
 \begin{equation}
     \frac{\partial \rho}{\partial t} +\nabla\cdot(\rho \vec{\varv})= 0
 \end{equation}
 \begin{equation}
     \frac{\partial 
(\rho \vec{\varv})}{\partial t} +\nabla\cdot\bigg[\rho \vec{\varv}\cdot\vec{\varv}+\bigg(P{+\frac{\vec{B}\cdot\vec{B}}{8\pi}}\bigg)\widetilde{I}-{\frac{\vec{B}\vec{B}}{8\pi}}-\widetilde{\tau}\bigg]= \rho \vec{g}
 \end{equation}
 
 \begin{equation}
    \frac{\partial E}{\partial t} +\nabla\cdot\bigg[\bigg(E+P+\frac{\vec{B}\cdot\vec{B}}{8\pi}\bigg)\vec{\varv}-\frac{(\vec{\varv}\cdot\vec{B})\vec{B}}{4\pi}\bigg]
    =\rho\vec{g}\cdot\vec{\varv}
 \end{equation}

 \begin{equation}
     {\frac{\partial \vec{B}}{\partial t} +\nabla\times(\vec{B}\times\vec{ \varv}+\eta_m\vec{J})= 0}
 \end{equation}
The above equations are continuity equation, momentum equation, energy equation and induction equation respectively. The symbols have their usual meaning: $\rho$ and $\vec{\varv}$ are the matter density and velocity, $P$ is the pressure and $B$ is the magnetic field. The viscous stress tensor is,
\begin{equation}
\widetilde{\tau}=\eta\left[(\nabla\vec{\varv})+(\nabla\vec{\varv})^T-\frac{2}{3}(\nabla\cdot\vec{\varv})\widetilde{I}\right]
\end{equation}
with the dynamic viscosity $\eta=\rho\nu_{\mathrm v}$, where $\nu_{\mathrm v}$ is the kinematic viscosity. This is computed using \citet{SS73} alpha parametrization for the kinematic viscosity $ \nu_{\mathrm v}= 2\alpha_\mathrm{v}C_\mathrm{s}^2/ (3\Omega_{\mathrm{K}\star})$ where $C_\mathrm{s}$ is the local sound speed. In the magnetic case, we assume that the magnetic diffusivity $\nu_{\mathrm m}$ is treated in a way analagous to kinematic viscosity: $\nu_{\mathrm m}= \alpha_{\mathrm m} C_\mathrm{s}^2/ \Omega_{\mathrm{K}\star}= 3 \alpha_{\mathrm m}\nu_{\mathrm v}/(2\alpha_{\mathrm v})$  \citep{ZF09}. Then in the cgs system of units the resistivity is $\eta_{\mathrm m}=4\pi\nu_{\mathrm m}$. While our results are obtained within resistive MHD with the resistivity explicitly present in the induction equation, we have excluded the dissipative viscous and resistive fluxes from the energy equation. We assume that all the produced heat is instantly radiated away locally, as described in \citet{cem19}, and in keeping with the thin disc approximation.

We perform a parameter study by changing the viscosity parameter $\alpha_{\mathrm v}$ and stellar rotation rate $\Omega_\star$ in hydrodynamic simulations, and add the variation of the resistivity parameter $\alpha_{\mathrm m}$ in magneto-hydrodynamic simulation. Both the viscosity and resistivity  parameters are varied from 0 to 1. Viscosity parameter describes the strength of the viscous torque, that allows the disc to accrete. The resistivity parameter $\alpha_{\mathrm m}$ allows coupling of the stellar magnetic field with the disc material. The combined effect of those two parameters is described by the magnetic Prandtl number:
\begin{equation}
    P_\mathrm{m}= \frac{2}{3}\frac{\alpha_\mathrm{v}}{\alpha_\mathrm{m}} .
\end{equation}
In MHD simulations with a similar setup \citep{ZF09, cem19} the viscosity parameter $\alpha_{\mathrm {v}}$ has been set to unity, constraining $P_\mathrm{m}$ to be above 0.66. Our simulations here are an extension of those simulations into a complete parameter space in the $\alpha_{\mathrm v}$.

Each simulation in our parameter study is run until a quasi-stationary state is reached. In practice, this means that we identify an interval after the relaxation in the simulations, in which for at least a period of 10 stellar rotations there is no large variation in the mass accretion and angular momentum fluxes in the simulation.  We scale $\Omega_\star$ with $\Omega_\mathrm{br}\equiv (GM_\star/R^3_\star)^{1/3}$, the Keplerian angular velocity at $r=R_\star$  (``breakup'' rotation rate, i.e., equatorial mass shedding limit). Time is given in units of stellar rotation periods $P_\star =2\mathrm{\pi}/\Omega_\star$.

\section{Results}   
\subsection{Backflow in HD disc}   
We obtain purely hydrodynamics solution using the HD module in PLUTO. One such simulation has been presented in Fig.~\ref{fullpihd}. We produce results for different values of viscosity parameter $\alpha_{\mathrm v}$ (Fig.~\ref{stag}) and for different stellar rotation rates $\Omega_\star$.  For the simulations with $\alpha_{\mathrm v} < 0.6$, after initial relaxation a stable backflow appears close to the equatorial plane of the computational domain. The radial velocity component in the disc points away from the star outside the stagnation radius, where the velocities turn away from the star, to the outer end of the computational domain. The smaller the value of $\alpha_{\mathrm v}$, the larger the portion of the disc is dominated by the backflow, as shown in Fig.~\ref{stag}. 

When $\alpha_{\mathrm v}>0.7$ there is no backflow in the disc, hence the radial velocity vectors point towards the star across the entire height of the disc. While this is in agreement with the $\alpha_{\mathrm {cr}}$ value of the KK solution (our Eq.~\ref{alfcrit}), that result was based on an assumption of steady flow. In our simulations, we do not assume {\it a priori} that the flow is steady. This is a significant distinction, as we will see in the discussion of the MHD results, which includes cases of intermittent backflow. We can therefore state one of the conclusions of our HD work: steady (i.e. quasi-stationary) solutions exist for the nonmagnetized $\alpha$-disc for values of $\alpha_{\mathrm v}$ both smaller and larger than $\alpha_{\mathrm {cr}}$. In the former case there is steady backflow, in the latter there is no backflow.

In the bottom panel of Fig.~\ref{hd} we present a snapshot from our simulation with backflow after a quasi stationary state is reached (we consider it quasi-stationary when the mass flux is almost constant). We chose a particular instant in time, as indicated by vertical line in the top right panel of Fig.~\ref{hd}. In the top left panel of Fig.~\ref{hd} we show that the radial velocities have different magnitude across different heights of the disc. The stagnation radius has been marked as the position where the radial velocity vectors change direction in the disc midplane. 
The inflow velocity at the disc surface is larger in magnitude than the backflow velocity at the disc midplane. In Fig.~\ref{vtheta} the negative of the latitudinal velocity component ($\varv_{\theta}$) has been shown as a function of ${\theta}$ at three different stellar radii from one of our simulations. Close to the midplane $(\theta=1.6)$ it is zero, but a little above the midplane  $-\varv_{\theta}$  has a slightly positive value. Thus, the fluid flows away from the midplane at the bottom of the backflow region, in agreement with the KK solution, and as expected from the simulations cited in Section~\ref{intro}.
We determine the stagnation radius from the plot of radial velocity at different distances from the star. The radial velocity vector along the disc midplane changes its sign at that radius, as shown in Fig.~\ref{hd}. We find that stagnation radius is a function of $\alpha_\mathrm{v}$, as presented in Fig.~\ref{FigVibStab}. It follows the same trend as predicted by the analytical model in \citet{KK00}. We obtained similar results for all stellar rotation rates.

    \begin{table*}
\centering
 \begin{tabular}{||c c c c c c c c||} 
 \hline
 Simulation & $\alpha_{\mathrm {v}}$ & $\alpha_{\mathrm {m}}$ & $P_{\mathrm {m}}$ & Backflow  &$r_{\mathrm{tr}}$ & $r_{\mathrm{stag}}$ &Type\\ [0.5ex] 
 \hline\hline
S1 & 0.1 & 1.0 & 0.06 & Yes & $4.5 \pm0.5$& $6\pm1$ &steady \\
 \hline
S2 &  0.2 & 1.0 & 0.13 & Yes& $6\pm1$ & $7.5\pm1.5$ & steady \\
  \hline
S3 & 0.3 & 1.0 & 0.20 & Yes& $4.5 \pm0.5$ & $6\pm1$ & steady \\
  \hline
S4 & 0.2 & 0.6 & 0.22 & Yes& $4.5 \pm0.5$& $6\pm1$ & steady \\
\hline  
S5 & 0.4 & 1.0 & 0.26  & Yes & $4.5 \pm0.5$& $6\pm 2$ & steady\\
 \hline
S6 & 0.5 & 1.0&  0.30 & Yes& $4.5 \pm0.5$ & $6\pm1$ & steady \\
  \hline
S7 & 0.6 & 1.0  &  0.40 & Yes& $4.5 \pm0.5$ & $6\pm1$ & steady \\
  \hline 
S8 & 0.4 & 0.6 & 0.44 & Yes& $4.5 \pm0.5$ & $6\pm1$ & steady \\
\hline
S9 & 0.7 & 1.0 &  0.46 & Yes& $3$ to $5^*$ & $\approx r_{\mathrm{tr}} + 1.5$ & intermittent$^*$\\
 \hline  
S10 & 0.8 & 1.0 & 0.53 & Yes& $3$ to $5^*$ &  $\approx r_{\mathrm{tr}} + 1.5$ &  intermittent$^*$\\
 \hline
S11 & 0.1 & 0.1 & 0.66 & No& - & -- & --\\ 
 \hline
 S12 & 0.4 & 0.4 & 0.66 & No& - & -- & -- \\
\hline
S13 & 1.0 & 1.0 & 0.66 & No & -& -- & -- \\
\hline 
 S14 & 1.0 & 0.4 & 1.65 & No& - & -- & -- \\
\hline
S15 & 0.4 & 0.1 & 2.64 & No& -& -- & -- \\
\hline  
S16 & 1.0 & 0.1 & 6.60 & No& - & -- & --\\   
\hline
 \end{tabular}
 \caption{\label{Table 1}List of parameters and results in our simulations: viscosity and resistivity coefficients $\alpha_{\mathrm {v}}$ and $\alpha_{\mathrm m}$, the magnetic Prandtl number $P_{\mathrm m}=2\alpha_{\mathrm v}/(3\alpha_{\mathrm m})$, presence of the backflow, truncation and stagnation radius in units of $R_*$, and type of backflow for a strongly magnetized, rotating star. $B_\star=1000~G$ at the stellar equator, and $\Omega_\star=0.1\,\Omega_\mathrm{br}$ with $r_\mathrm{co}= 4.6\, R_\star$ in all the cases. 
  $^*${\sl Note:} In the intermittent case, both the inner edge of the disc and the stagnation radius move in and out in concert, with values depending on the phase of the quasiperiodic midplane flow.}
\end{table*}

\subsection{Backflow in MHD disc}
We obtain MHD solutions in a resistive accretion disc for different values of $\alpha_{\mathrm {v}}$ and $\alpha_{\mathrm {m}}$ for a slowly rotating star with a strong magnetic dipole field of $1000\,$G at the stellar equator (magnetic dipole of $\sim 1.4 \times10^{36}\,\mathrm{G\,cm^3}$). 
A snapshot in our simulation after a quasi stationary state is reached is shown in Fig.~\ref{Mhd}. In the top-right panel we present the mass accretion rate as a function of time and the vertical line in the plot indicates the snapshot in the bottom panel. The accretion disc is truncated at a few stellar radii due to magnetic pressure as shown in the middle and bottom panels of Fig.~\ref{Mhd}. The accretion flow is channelled into a funnel where the flow follows the magnetic field lines. 

On theoretical grounds, the disc should be truncated where the magnetic field pressure balances the ram pressure of the accreting material. For practical reasons, we define the truncation radius, $r_\mathrm{tr}$ as the radius of the inner edge of the accretion disc and determine it from the density and velocity contours, rather than the theoretical formula for the magnetospheric truncation radius. Coincidentally, for the stellar parameters and mass accretion rate chosen for this study, the co-rotation radius, $r_\mathrm{co}=4.6 \, R_\star$ is very close to the truncation radius in all cases.
In Fig.~\ref{Mhd}, middle panel, we can distinguish that there is accretion along the magnetic field lines in both upper and lower part of the inner disc, and this is accompanied by a decrease in the fluid density. Close to the foot of accretion column in the middle middle panel, the (already decreasing) density suffers a sudden and profound dip, and that is where we mark the truncation radius ($r_\mathrm{tr}$) in the disc.  

Backflow appears in the disc midplane only for some combinations of $\alpha_{\mathrm {v}}$ and $\alpha_\mathrm {m}$ parameters\footnote{Because of numerical reasons, in some cases our simulation stops after a few stellar rotations. We only report successful runs.}. We list these parameters in 
Table~\ref{Table 1}. Unlike in HD cases, in MHD simulations we find a circulation pattern at the stagnation radius very close to the star as shown by the streamlines in the middle panel of Fig.~\ref{Mhd}. Backflow is carried until the end of the disc. In the upper region of the disc (i.e. at high altitudes above the midplane) the accretion flow is towards the central star.
To distinguish the inflow and backflow region more clearly, we refer to the radial velocity plot as indicated in the bottom panel of Fig.~\ref{Mhd}. One can clearly infer the stagnation radius from this plot. Radial velocity has different magnitude at different disc heights, as explained in the top-left panel of Fig.~\ref{Mhd}.

We find that for a fixed coefficient of viscosity $\alpha_\mathrm {v}$ lower values of the resistive parameter $\alpha_\mathrm {m}$ restrict backflow in the disc. Conversely, when $\alpha_\mathrm {m}$ attains large values, in the MHD simulations backflow can occur even for higher values of critical viscosity than obtained in HD case.
In fact, as we demonstrate in Table~\ref{Table 1}, the presence of backflow dependends on magnetic Prandtl number alone, and not on the values of the viscous and resisitive coefficients separately. Above a critical value of magnetic Prandtl number, which is about $P_{\mathrm {m}}^\mathrm{crit}\sim 0.6$, there is no backflow in the disc in our simulation. In the case of slowly rotating stars, the stagnation radius in the steady MHD disc appears to be independent of alpha viscosity (Fig.~\ref{FigVibStab}), in marked contrast to the HD disc. We indicate the stagnation radius as a function of $\alpha$ in Fig.~\ref{FigVibStab}. The blue points with error bars are from intermittent backflow cases. We fit the steady stagnation radius in MHD simulations with a constant (horizontal solid red line).  However, in intermittent flow the stagnation radius is slightly larger (blue error bars).

When we do attain backflow, it can be of two categories: a steady  flow as shown in Fig.~\ref{Mhd} or an intermittent flow as shown in Fig.~\ref{int}. A steady  backflow is one where backflow stays stationary when we evolve the disc in time. An intermittent backflow appears as a function of time where backflow persist for certain time intervals and disappears in others.  
Interestingly, in the few cases of intermittent flow that we investigated, it occurs when $\alpha_\mathrm {v}$ exceeds the critical value for backflow in HD. Future investigations will show whether this breaks the dependence on the Prandtl number alone. The quasiperiodic change from backflow to inflow in the midplane region is accompanied by a corresponding change in the value of the truncation radius (see \href{https://drive.google.com/file/d/185n7UgvlSYvxOUsgJdR4lOW9qqKzwEqu/view?usp=sharing}{movie}\footnote{\url{https://tinyurl.com/rmbackflow}}. When present, the stagnation radius is always at about the same distance from the inner edge of the disc, whether in steady or intermittent backflow, with $r_\mathrm{stag} \approx r_\mathrm{tr} + 1.5\, R_\star$. An inspection of the detailed time evolution of the disc suggests that the intermittent backflow is achieved when the squeezed magnetic field of the central star pushes back on the inner flow, resulting in the backflow region being established throughout the disc (i.e., for all radii). After some time the inflow that would be usual in the HD case for these high values of $\alpha_\mathrm {v}$) is reestablished in the disc. The intermittent midplane backflow is accompanied by ``field inflation'', i.e. closed field lines anchored in the disc get pushed out to larger radii (in the backflow stage) and pushed back to the inner parts of the system (in the inflow stage). In this way the backflow occuring at the bottom of the disc (close to the midplane) directly affects conditions at the surface of the disc, and even outside it. In principle, this coupling could lead to additional observational signatures of this inner sloshing mode.

This MHD behaviour differs from the hydrodynamic case, where the stagnation radius is a strong function of the viscosity parameter. In the MHD case, the steady backflow region seems to be present throughout most of the disc, and with a constant value of the stagnation radius, for Prandtl numbers $P_{\mathrm {m}} < 0.45$. For $0.45 < P_{\mathrm {m}} < P_{\mathrm {m}}^\mathrm{crit} < 0.66$ the backflow is intermittent, with a slightly increased value of the stagnation radius. For values larger than $P_{\mathrm {m}}^\mathrm{crit}$ there is no backflow. Thus, in marked contrast to the purely HD case, there never seems to be an MHD case where the backflow would occur in the outer parts of the disc alone.

\subsection{Mass flow rates}
The mass flow rate crossing the whole domain is the integral of the mass flux inside and outside of the disc. It is given by\\ 
\begin{equation}
    \dot{M}= - 4\pi r^2\int_0^{\pi/2} \rho \varv_{r} \sin{\theta} \,d \theta \ .
    \label{massflow}
\end{equation}
The limit of the integration is suitable for our $(0,\pi/2)$ setup. In the disc part which has both inflow and backflow part, we can split the mass accretion rate into a difference between the inflowing mass flux and the outflowing one.
In Fig.~\ref{massflux} we present the  total mass flux, and the mass flux in the backflow, as a function of radius and as a function of time for an MHD simulation. We use the usual sign convention that a positive mass flow rate corresponds to accretion. Hence, this quantity is negative in the backflow. The mass flux as a function of radius stays almost constant and there is very little variation in mass flux as a function of time. This indicates  that the simulation is quasi-stationary.

\section{Discussion}
In this Section we review arguments for and against the reality of backflow in accretion discs.
When starting this project we aimed to obtain a self consistent MHD solution of star-disc magnetosphere interaction, as in \citet{ZF09}. We found backflow in the accretion disc while spanning the parameter space in the simulations. 
We have focused the presentation in this paper on backflow in the MHD accretion disc, as its presence had not been anticipated. 

Backflow is well established analytically and numerically in HD discs, as well as in geometrically thick flows driven by MRI (see Section~\ref{intro}).
However, as mentioned above, a number of studies do not pay much attention to backflow and reason it away as an artefact of alpha-prescription. 
Indeed, a number of numerical and analytical studies are based on models where angular momentum transport results from conventional alpha viscosity \citet{SS73}, while  in many scenarios angular momentum transport is more likely to be caused by MRI \citep[magneto rotational instability,][]{bablu}. However, MRI solutions do support the \citet{SS73} $\alpha$-disc prescription, at-least in the leading term of stress-energy tensor \citep{hirose}, and although the form of the stress tensor is different in MRI and viscous discs, with some hindsight we could have then expected the main features of the viscous disc solution to perhaps be present also in the MHD disc, whether one with an effective viscosity (as we found in our $\alpha$-disc prescription MHD simulations) or a disc explicitly driven by MRI turbulence (\citealp[as in the recent MRI simulations of][where backflow has also been reported]{bm}).

Some numerical simulations of turbulent disc within the framework of MRI, such as global MHD simulations of \citet{flock}; \citet{fromang}; \citet{suzuki}, do not observe backflow in the disc midplane. 
In \citet{jack}, the author makes a comparison between a viscous prescription and a MHD prescription and indicates that the anisotropy of the stress tensor $T_{ij}$ can lead to backflow in the midplane. The author also comments on the physical reality of these ``outflows'' as controversial. However, \citet{bm} have now clearly found  an equatorial backflow in what is essentially a dense thin disc with an elevated accreting region. \citet{bm} suggests that the polar stress  $T_{\theta \phi}$ behaves differently in the region close to the disc midplane as compared to higher altitudes. $T_{\theta \phi}$ points towards the disc midplane which could lead to angular momentum deposition on the disc midplane, causing backflows.

Backflow in the accretion disc can also be inferred from various observations. There has been evidence of backflow (outward transport of material) occurring at the protoplanetary disc \citet{keller,cia,TakuLin}. Usually crystalline dust material is formed in the inner warm region of the accretion disc, closer to the central object. Analysis of the material ejected from comet Wild 2 \citet{bro} reveals the presence of high temperature minerals. These minerals are usually accreted by the comets from the outer region of protoplanetary discs. Hence, backflow may explain the transport of a crystaline dust material to the outer cold region of the protoplanetary disc.

\section{Conclusions}
We conclude this paper with the following summary and comments.
   \begin{enumerate}
      \item We present resistive magnetohydrodynamics simulation of thin $\alpha$-discs in accretion onto magnetized and nonmagnetized rotating stars for different values of $\alpha_{\mathrm v}$ and $\alpha_\mathrm m$ parameters.
      
     \item For purely hydrodynamic simulations of the thin $\alpha$-disc we find backflow in the midplane of the accretion disc for $\alpha_{\mathrm v}<\alpha_{\mathrm {cr}}\approx 0.7$ in our simulations, in agreement with the analytic work of KK (in spite of different inner boundary conditions).
         
         \item Quasi-stationary solutions exist for the nonmagnetized (HD) $\alpha$-disc for values of $\alpha_{\mathrm v}$ both smaller and larger than $\alpha_{\mathrm {cr}}$. In the former case there is steady backflow, in the latter there is no backflow.
         
      \item We find that the stagnation radius in the midplane, which separates the (inner) region of inflow from (the outer) region of backflow, follows the dependence on $\alpha_{\mathrm v}$ predicted by the analytical model \citet{KK00,2002RG}.
      
      \item We find that the character of backflow (steady or not, present or absent) in the resistive MHD simulations depends on the magnetic Prandtl number, $P_\mathrm m$. In Table 1, when  $P_\mathrm m<0.6$ there is a backflow. As the critical value of $P_\mathrm m=0.6$ is approached the backflow in the disc becomes intermittent.

      \item In the MHD case, the backflow occuring deep inside the disc is coupled to the conditions outside the disc through magnetic field lines threading the disc. Potentially, this could lead to directly observable manifestations of backflow, particularly in the intermittent case.
      
       \item The intermittent backflow manifests itself at any given radius in extended periods of backflow interspersed with (perhaps shorter, see Fig.~\ref{int}) periods of inflow. This may manifest itself as a QPO (quasi-periodic oscillation), if transmitted to some observable quantity. The corresponding quasi-period is rather long, on the order of 100 stellar periods, in our simulations (or $\sim 30$ orbital periods in the inner disc at about 10 stellar radii).
       
       \item In the presence of strong magnetic field and high resistivity (so that $P_\mathrm m$ is below its critical value), we obtain backflow in MHD disc for even higher values of viscous parameter than critical $\alpha_\mathrm v$ in the purely HD case.
     
      \item In contrast with the HD case, where the stagnation radius is strongly dependent on $\alpha_{\mathrm v}$, in the MHD case the stagnation radius has the same value for all values of the viscous and diffusion coefficients, as long as $P_\mathrm m<0.45$ (at least for the stellar rotation rate and magnetic dipole reported in Table 1). Whenever the midplane backflow is present in the MHD simulations, it is present essentially throughout the disc, extending all the way inwards to the transition zone, where matter is channeled into the accretion funnel at higher altitudes in the disc (Fig.~\ref{Mhd}). In this case the stagnation circle separates the region of backflow from the region where matter is lifted even from the midplane regions into the accretion funnel connected to the circumpolar regions of the stellar surface.
         
   \end{enumerate}
   
 In the future we would like to analyze the quasi-periodic oscillation of the midplane fluid in the intermittent backflow. Also, a more extensive exploration of the parameter space is needed in the resistive MHD simulations, e.g. different stellar rotations, different magnetic field strengths etc., if only to confirm our finding on the Prandtl number dependence of the backflow character.
 We also leave for the future a discussion of the influence of backflow on torques in the star-disc magnetosphere.  We also intend to investigate the effects of backflow on the parameter ``f'' which is the ratio of outward viscous flux of angular momentum from the inner boundary to the inward advected flux of angular momentum there in reference to \citet{jep}, as well as effects of backflow on the positioning of material in the protoplanetary environment.
 
 The main caveat in our work here is that our simulations are not fully 3D, but axisymmetric (2.5 D).
\section*{Acknowledgements}This project was funded by a Polish NCN grant No. 2019/33/B/ST9/01564. In addition, M\v{C} acknowledges ANR Toupies grant under which he developed the setup for star-disc simulations while in CEA, Saclay, and also a collaboration with Croatian project STARDUST through HRZZ grant IP-2014-09-8656. His work in Shanghai Observatory during a part of work on the project was funded by CAS President’s International Fellowship for Visiting Scientists (grant No. 2020VMC0002), and in Opava by the ESF projects No. $\mathrm{CZ.02.2.69/0.0/0.0/18\_054/0014696}$. We thank ASIAA (PL and XL clusters) in Taipei, Taiwan and CAMK (CHUCK cluster) in Warsaw, Poland, for access to Linux computer clusters used for high-performance computations. We thank the {\sc pluto} team for the possibility to use the code.
\section*{Data availability}
The simulation data associated with this article will be made available by the corresponding author upon reasonable request.
\bibliographystyle{mnras}
\bibliography{ruchibackflow} 



\bsp	
\label{lastpage}
\end{document}